\documentclass[aps,prc,twocolumn,showpacs,superscriptaddress]{revtex4}
\usepackage[latin1]{inputenc}
\usepackage{bm}
\usepackage{epsfig}
\usepackage{amsthm}
\usepackage{amsfonts}
\usepackage{float}
\usepackage{amsmath,amssymb}
\usepackage{graphicx}
\usepackage{dcolumn}
\usepackage{bm}
\def\nn{\nonumber}
\newcommand{\be}{\begin{equation}}
\newcommand{\ee}{\end{equation}}
\newcommand{\bea}{\begin{eqnarray}}
\newcommand{\eea}{\end{eqnarray}}

\newcommand{\ep}{\epsilon}

\newcommand{\om}{\omega}

\newcommand{\ov}{\overline}

\newcommand{\qs}{q \!\!\! /}
\newcommand{\ks}{k \!\!\! /}

\newcommand{\vk}{\vec k}

\newcommand{\vq}{\vec q} 
\newcommand{\vl}{\vec l}

\newcommand{\del}{\partial}

\newcommand{\unit}{1\!\!1}
\begin{document}
\title{Electrical conductivity of hadronic matter from different possible mesonic and baryonic thermal fluctuations}
\author{Sabyasachi Ghosh}
\email{sabyaphy@gmail.com}
\affiliation{Department of Physics, University of Calcutta, 92, A. P. C. R
oad, Kolkata - 700009, India}
\begin{abstract}
Electromagnetic current-current correlators in pionic and nucleonic medium
have been evaluated in the static limit to obtain electrical conductivities for pion
and nucleon components respectively, where former decreases and latter one increases
with the variation of temperature $T$ and baryon chemical potential $\mu_N$. Therefore, total
electrical conductivity of pion and nucleon system exhibits a valley structure in the 
$T$-$\mu_N$ plane. To get non-divergent and finite values of correlators, finite thermal
widths of medium constituents, pion and nucleon have been considered, where these thermal
widths have been determined from the in-medium scattering probabilities
of pion and nucleon with other mesonic and baryonic resonances, based on effective hadronic
model. At $\mu_N=0$, the results of present work are more or less agrees with the results of
earlier works and its finite $\mu_N$ extension show a decreasing nature of electrical
conductivity for hadronic medium during spanning from freeze out line to quark-hadron
transition line in $T$-$\mu_N$ plane.
\end{abstract}
\maketitle
%
%
%
\section{Introduction}
\label{sec:intro}
The electromagnetic current-current correlator at finite temperature 
is one of the very important quantity to characterize the medium,
produced in high energy heavy ion collisions. The explicit dynamical
structure of this quantity for hadronic matter (HM) is directly linked with the in-medium
spectral function of neutral vector mesons and also with the thermal
dilepton and photon yields from HM sources, whereas its static limit provide the 
estimation of an important transport coefficients like electrical
conductivity ($\sigma$) of the HM. According to 
recent reviews~\cite{Rapp_rev,G_rev}, the effective field theoretical calculations 
of hadrons at finite temperature are
very successful to describe the low mass dimuon enhancement measured by the NA60
collaboration~\cite{SPS}. This low mass enhancement also get boost from the quark 
matter (QM) sources, which has been calculated by using Hard Thermal Loop (HTL) 
technique in 
Ref.~\cite{HTL} (see also Ref.~\cite{Sarbani} for effective QCD model calculation).
Therefore, it will be very interesting and 
phenomenologically important to know
the static limit estimation of the dynamical structure of current-current correlator
by calculating $\sigma$ of hadronic medium in the frame work
of effective hadronic model, which is basically attempted by this present work.

The event by event analysis~\cite{Sokov} in relativistic heavy ion collisions 
indicates about the possibility of generation of a high strength electric ($E$) and 
magnetic ($B$) fields in the medium. For example, in the 
relativistic heavy ion collider (RHIC) experiment, their approximate
values are $eB\approx m_\pi^2\approx 10^{18} G$ and 
$eE\approx m_\pi^2\approx 10^{21}V/cm$~\cite{Tuchin}. Although a particular 
magnetic field component becomes only non-zero in the average 
scenario~\cite{Sokov,Tuchin}.
The time evolution of this average magnetic field~\cite{Tuchin} depends on
the $\sigma$ of the expanding medium, produced in 
heavy ion collisions, which demands that we should have some good idea
on numerical values of this $\sigma$.

In Ref.~\cite{Akamatsu}, the electrical conductivity or electric
charge diffusion coefficient of evolving medium is used as input
to explain the low mass dilepton enhancement, observed experimentally
by PHENIX collaboration at RHIC. Whereas, Yin~\cite{Yin} have shown
that the electrical conductivity of quark-gluon-plasma (QGP) plays
important role to regulate the soft photon production via 
realistic hydrodynamics simulation. Besides these indirect estimation
of electrical conductivity of QGP, it can directly be extracted from 
charge dependent direct flow parameters in asymmetric heavy-ion
(Au+Cu) collisions~\cite{Hirano}. Along with these phenomenological
searching, different microscopic calculations for $\sigma$ of quark 
~\cite{Cassing,Marty,Puglisi,Greif,PKS,Finazzo} and
hadronic phase~\cite{Lee,Nicola_PRD,Nicola,Greif2} have been done, although the results
of Cassing et al~\cite{Cassing} in the model of PHSD (parton hadron string dynamics)
and the NJL (Nambu-Jona-Lasino) model results of Marty et al.~\cite{Marty} have 
covered $\sigma$ estimation for the temperature domain of both quark and hadronic matter.
On this problems, a large number of
Lattice QCD calculations have been done~\cite{LQCD_Ding,LQCD_Arts_2007,LQCD_Buividovich,
LQCD_Burnier,LQCD_Gupta,LQCD_Barndt,LQCD_Amato}, where their estimations 
cover a large numerical band (see table~\ref{tab}, addressed in result section).
Now, from the calculations~\cite{Lee,Nicola_PRD,Nicola,Greif2} in the 
hadronic temperature domain, we see that the results
of Ref.~\cite{Lee} and Refs.~\cite{Nicola_PRD,Nicola,Greif2} show
completely opposite nature of temperature ($T$) dependence of $\sigma$.
If we considered the results of Lee et al.~\cite{Lee} as an exceptional,
almost all of the earlier works~\cite{Cassing,Marty,Puglisi,Greif,PKS,Finazzo,
Nicola_PRD,Nicola,Greif2,LQCD_Amato} indicates that $\sigma/T$ decreases in
hadronic temperature domain~\cite{Cassing,Marty,Nicola_PRD,Nicola,Greif2} 
and increases in the temperature domain of 
quark phase~\cite{Cassing,Marty,Puglisi,Greif,PKS,Finazzo,LQCD_Amato}. 
Their numerical values are located within the order - $\sigma/T\approx 10^{-3}$ to $10^{-2}$
for hadronic phase and $\sigma/T\approx 10^{-3}$ to $10^{-1}$ for quark phase.
These information from earlier studies indicate that the numerical strength as well as
the nature of $\sigma(T)$ both are not very settle issue till now.  

In this context, the present investigation is similar kind of microscopic calculations
for $\sigma$, which is expected
to converge and update our understanding of $\sigma(T)$. Considering pion and nucleon as abundant
constituents of hadronic matter, we have calculated their electromagnetic current-current
correlators at finite temperature, whose static limit give the estimation of $\sigma$
for the respective components. As an interaction part, the effective hadronic 
Lagrangian densities have been used to calculate
the in-medium scattering probabilities of pion and nucleon with other mesonic
and baryonic resonances, present in the hadronic medium.
Extending our investigations for 
finite nucleon or baryon chemical potential $\mu_N$, the present results
provide the 
estimation of
$\sigma$ in $T$-$\mu_N$ domain of hadronic matter. 

The basic formalism of $\sigma$ is addressed in 
the Sec.~\ref{sec:form}, where we will see that 
the non-divergent values of current-current correlator
are mainly regulated by the thermal widths of medium components, which are
calculated and briefly described in Sec.~\ref{sec:width}. 
Calculations of different loop diagrams are classified in three subsections.
After it, the numerical discussions have been
addressed in the result section (Sec.~\ref{sec:num}), which is followed by a summary in
Sec.~\ref{sec:concl}.

\section{Formalism of electrical conductivity}
\label{sec:form}
Owing to the famous Kubo formula~\cite{Zubarev,Kubo},
the electrical conductivity in momentum
space can be expressed in terms of spectral density
of current current correlator as~\cite{Nicola} 
\be
\sigma=\frac{1}{6}\lim_{q_0,\vq \rightarrow 0}\frac{A_\sigma(q_0,\vq)}{q_0}
\label{sigma_Nicola}
\ee
where $A_\sigma(q_0,\vq)=\int d^4x e^{iq\cdot x}\langle[J^{\rm EM}_{i}(x),J^{i}_{EM}(0)]\rangle_\beta$
with
$\langle ..\rangle_\beta$
denotes the thermodynamical ensemble average.

In real-time thermal field-theory (RTF), any two point function at finite temperature
always gives a $2\times 2$ matrix structure. Hence, the thermal correlator
of electromagnetic current ($J^{\rm EM}_{\mu}(x)$) will be
\be
\Pi^{ab}(q)=i\int d^4x e^{iqx}\langle T_c J^{\rm EM}_{\mu}(x) J_{\rm EM}^{\mu}(0)\rangle^{ab}_\beta~,
\label{pi_ab}
\ee
where $T_c$ denotes the time ordering with respect to a 
symmetric contour in the complex time plane. Because of the contour,
we get four possible set of two points and Therefore we get $2\times 2$ 
matrix structure of two point function. The superscripts 
$a, b (=1,2)$ in Eq.~(\ref{pi_ab}) represent the (thermal) indices
of the matrix.
%
%
Retarded part of correlator $\Pi^R(q)$ and its corresponding
spectral density $A_\sigma(q)$ can be extracted from 
11-component $\Pi^{11}(q)$ by using the relation
\be
A_\sigma(q)=2{\rm Im}\Pi^R(q)=2{\rm tanh}(\frac{\beta q_0}{2})
{\rm Im}\Pi_{11}(q)~.
\label{A_piR}
\ee
Using this relation (\ref{A_piR}), 
the Eq.~(\ref{sigma_Nicola}) can alternatively be expressed as
\bea
\sigma&=&\frac{1}{3}\lim_{q_0,\vq \rightarrow 0}
\frac{{\rm Im}\Pi^R(q_0,\vq)}{q_0}
\nn\\
&=&\frac{1}{3}\lim_{q_0,\vq \rightarrow 0}
\frac{{\rm tanh}(\frac{\beta q_0}{2}){\rm Im}\Pi^{11}(q_0,\vq)}{q_0}~.
\label{el_Im_R}
\eea

Since pion and nucleon constituents are our
matter of interest, so we should focus
on their electromagnetic currents:
\bea
J^\mu_{\pi}&=&e\phi_\pi(\del^\mu\phi_\pi)
\nn\\
{\rm and}~J^\mu_{N}&=&e{\ov\psi}_N\gamma^\mu\psi_N~,
\label{current}
\eea 
which are electromagnetically coupled with photon via
interaction (QED) Lagrangian density
\be
{\cal L}=-(J^\mu_\pi + J^\mu_N)A_\mu~.
\ee
Since ($\phi_{\pi^+}$, $\phi_{\pi^-}$) from  pion triplet 
($\phi_{\pi^+}$, $\phi_{\pi^-}$, $\phi_{\pi^0}$) and proton 
($\psi_p$) from nucleon doublet ($\psi_p$, $\psi_n$)
have non-zero electric charges, so
we have to keep in mind about relevant isospin factors $I^e_\pi=2$
and $I^e_N=1$, which should be multiplied during our calculations.

\begin{figure}
\begin{center}
\includegraphics[scale=0.52]{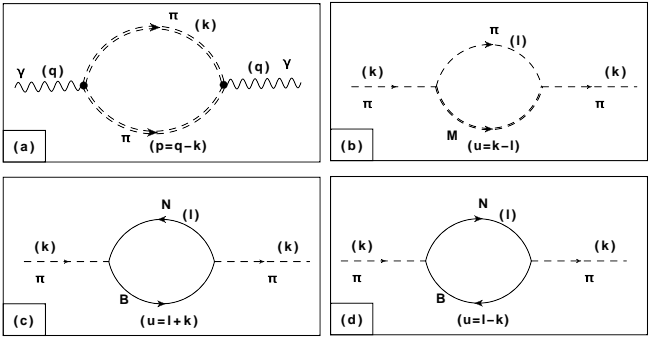}
\caption{The diagram (a) is a schematic one-loop representation of 
electromagnetic current-current correlator
for the medium with pionic constituents. The external photon lines
are coupled with double dashed internal lines of pions, which have some finite thermal
width. Thermal width of pion can be derived from its self-energy diagrams (b), (c) 
and (d), where (b) represents pion self-energy for mesonic ($\pi M$) loops,
whereas diagrams (c) and (d) are direct and cross diagrams of pion self-energy 
for $NB$ loops.} 
\label{El_pi_N}
\end{center}
\end{figure}

\begin{figure}
\begin{center}
\includegraphics[scale=0.52]{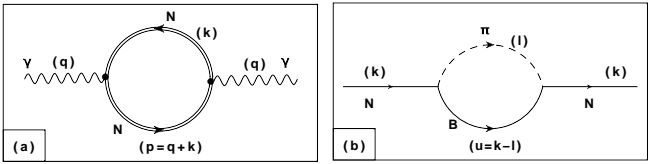}
\caption{The diagram (a) is a schematic one-loop representation of 
electromagnetic current-current correlator
for the medium with nucleonic constituents. Similar to double dashed lines 
of pions in Fig.~(\ref{El_pi_N}), here double solid lines of nucleon
indicates that they have finite thermal width, which can be obtained from 
the nucleon self-energy diagram (b) for $\pi B$ loops.} 
\label{El_N_piB}
\end{center}
\end{figure}

To calculate electrical conductivity of pionic ($\sigma_\pi$)
and nucleonic ($\sigma_N$) medium from their corresponding
spectral density or retarded part of correlator via Eq.~(\ref{el_Im_R}), 
let us start from
11-component of the $\Pi_{ab}$ matrix. 
The Wick contraction (see Appendix~\ref{cal_Nqk} ) of the
pion ($\phi_\pi$) and nucleon ($\psi_N$) fields give
one-loop diagrams of photon self-energy, which are shown 
in Fig.~\ref{El_pi_N}(a) and ~\ref{El_N_piB}(a) respectively. 
A general mathematical expression of these diagrams is
\be 
\Pi^{11}(q)=i e^2\int \frac{d^4k}{(2\pi)^4}N~ D^{11}(k)
D^{11}(p)~,
\label{self_eta}
\ee 
where $D^{11}(k)$ and $D^{11}(p)$ are the scalar parts of
propagators, appeared in RTF for 11-component; $p=q-k$ for
$\pi\pi$ loop in Fig.~\ref{El_pi_N}(a), $p=q+k$ for 
$NN$ loop in Fig.~\ref{El_N_piB}(a).
Multiplication of vertex part and numerator part of two propagators
build the term $N$.

In RTF, a general form of $D^{11}(k)$ for boson or fermion is
\bea
D^{11}(k)&=&\frac{-1}{k_0^2-\om_k^2+i\ep}+2\pi i\ep_k
F_k(k_0)\delta(k_0^2-\om_k^2)~,
\nn\\
&&~{\rm with}~ F_k(k_0)=n^+_k\theta(k_0)+
n^-_k\theta(-k_0)~,
\label{de11}
\eea
where $n^{\pm}_k(\om_k)=\frac{1}{e^{\beta(\om_k \mp \mu)} - \ep_k}$
are the thermal distribution functions and $\pm$ sign in 
the superscript of $n_k$ stand for particle and 
anti-particle respectively. Now when we proceed for special
cases - pion (boson) or nucleon (fermion) field, we have to
put
\bea
&&\ep_k=+1,~\mu=\mu_\pi=0~{\rm i.e.}~ n^+_k=n^-_k~,
\nn\\
&&\om_k=\om^{\pi}_k=(\vk^2+m_\pi^2)^{1/2}~{\rm for~pion}~,
\label{pi_case}
\eea 
\bea
&&\ep_k=-1,~\mu=\mu_N~({\rm nucleon~chemical~potential})~,
\nn\\
&&\om_k=\om^{N}_k=(\vk^2+m_N^2)^{1/2}~{\rm for~nucleon}~.
\label{N_case}
\eea 
However, for time being we will will continue our calculation
with the general form of $D^{11}$ from Eq.~(\ref{de11}) and
at latter stage, we will put these conditions (\ref{pi_case})
and (\ref{N_case}) in the general expression. 

After using (\ref{de11}) in Eq.~(\ref{self_eta}), if we do its $k_0$ integration
and put it in Eq.~(\ref{el_Im_R}), then we will get spectral
density of electromagnetic current-current correlator~\cite{G_IJMPA}:
\bea
A_\sigma(q)&=&e^2\int\frac{d^3k}{(2\pi)^3}
\frac{(-\pi)N}{4\om_k\om_p}[C_1\delta(q_0 -\om_k-\om_p)
\nn\\
&&+C_2\delta(q_0-\om_k+\om_p)+C_3\delta(q_0 +\om_k-\om_p)
\nn\\
&&+C_4\delta(q_0 +\om_k+\om_p)]~,
\label{Pi_LU}
\eea
where $\om_p=\om_p^{\pi}=\{(\vq -\vk)^2+m^2_{\pi}\}^{1/2}$ for pion field,
$\om_p=\om_p^{N}=\{(\vq +\vk)^2+m^2_{N}\}^{1/2}$ for nucleon field.
Here $N$ are space component of $N(q,k_0=\pm \om_k,\vk)$ (see Appendix ~\ref{cal_Nqk}):
\be
N= (-4)\{-\vk\cdot\vq + \vk^2\}~{\rm for}~\pi\pi~{\rm loop}~,
\label{N_qvkv_BB}
\ee
and
\bea
N&=&(-8)\{\vk\cdot\vq +\vk^2\} +4\vk\cdot\vq ~{\rm for}~NN~{\rm loop}~.
\label{N_qvkv_FF}
\eea
The statistical probabilities,
attached with four different delta functions, are
\bea
C_1&=& 1+n^{+}_k(\om_k)+n^+_p(q_0 - \om_k)~,
\nn\\
C_2&=&-n^{+}_k(\om_k)+n^-_p(-q_0+\om_k)~,
\nn\\
C_3&=&n^-_k(\om_k)-n^+_p(q_0+\om_k)~,
\nn\\
C_4&=&-1-n^-_k(\om_k)-n^-_p(-q_0-\om_k)~,~{\rm for}~\pi\pi~{\rm loop}~;
\nn\\
\eea 
and
\bea
C_1&=& -1+n^{-}_k(\om_k)+n^+_p(q_0 + \om_k)~,
\nn\\
C_2&=&-n^{-}_k(\om_k)+n^-_p(-q_0+\om_k)~,
\nn\\
C_3&=&n^+_k(\om_k)-n^+_p(q_0+\om_k)~,
\nn\\
C_4&=&1-n^+_k(\om_k)-n^-_p(-q_0-\om_k)~,~{\rm for}~NN~{\rm loop}~.
\nn\\
\eea
Four different delta functions are responsible for creating four 
different regions of branch cuts in $q_0$-axis, where $A_\sigma(q_0,\vq)$
or Im$\Pi^R(q_0,\vq)$ becomes non-zero.
These regions are
\bea
q_0&=&-\infty ~~~{\rm to}~~~ -\{\vq^2+4m_{\pi,N}^2\}^{1/2}~:~ {\rm unitary~ cut}~,
\nn\\
&=&\left.
\begin{array}{c}
 -|\vq| ~~~~~{\rm to}~~~~ ~0 \\
 ~~~0~~~~~{\rm to}~~~~ ~|\vq|
\end{array}
\right\} ~:~ {\rm Landau~ cut}~,
\nn\\
&=&\{\vq^2+4m_{\pi,N}^2\}^{1/2} ~~~{\rm to}~~~\infty ~:~ {\rm unitary~ cut}~.
\eea
Since electrical conductivity $\sigma$ is the limiting value
of $A_\sigma(q_0,\vq)$ or Im$\Pi^R(q_0,\vq)$ at $q_0, \vq\rightarrow 0$,
therefore we should focus on Landau cuts only.
Hence, using the Landau part of Eq.~(\ref{Pi_LU}) in Eq.~(\ref{sigma_Nicola}), we have
\bea
\sigma&=&\frac{e^2}{3}\lim_{q_0,\vq \rightarrow 0}\frac{1}{q_0}
\int\frac{d^3k}{(2\pi)^3}\frac{(-\pi)N}{4\om_k\om_p}
\{C_2\delta(q_0-\om_k+\om_p)
\nn\\
&&+C_3\delta(q_0+\om_k-\om_p)\}
\nn\\
&=&\frac{e^2}{3}\lim_{q_0,\vq \rightarrow 0}{\rm Im}
\left[\int\frac{d^3k}{(2\pi)^3}\frac{N}{4\om_k\om_p}\lim_{\Gamma \rightarrow 0}
\right.\nn\\
&&\left.
\left\{\frac{C_2/q_0}{(q_0-\om_k+\om_p)+i\Gamma}
+\frac{C_3/q_0}{(q_0+\om_k-\om_p)+i\Gamma}\right\}\right]~.
\nn\\
\label{el_G}
\eea
We will take finite value of $\Gamma$ in our further
calculations to get a non-divergent values of $\sigma$.
In Kubo approach, this traditional technique is widely
used to calculate different transport
coefficients like shear viscosity~\cite{Nicola,G_IJMPA}, 
electrical conductivity~\cite{Nicola_PRD}.
In this respect, this formalism
is very much close to quasi particle approximation.
The $\Gamma$ of medium constituents is basically their thermal width,
which is physically related with the probabilities of different
in-medium scattering. Inverse of $\Gamma$ measures the relaxation time
$\tau$, which is the average time of medium constituents
to reach their equilibrium conditions.

Next, applying the L'Hospital's rule in the Eq.~(\ref{el_G})
(see Appendix ~\ref{LHos}), 
we get a generalized expression
of electrical conductivity for bosonic ($\phi_\pi$) or fermionic ($\psi_N$) field:
\be
\sigma=\frac{\beta e^2}{3}\int\frac{d^3k}{(2\pi)^3}\frac{(-N^0)}{4\om_k^2\Gamma}
[n^-_k(1+\ep_kn^-_k)+n^+_k(1+\ep_kn^+_k)]~,
\label{el_last}
\ee
where 
\be
N^0=\lim_{q_0, \vq\rightarrow 0}N(k_0=\pm\om_k,\vk,q)~.
\label{N_0}
\ee
Depending upon the sign of $\ep_k$, the statistical probability
becomes Bose enhanced ($\ep_k=+1$ for bosonic field) or Pauli blocked
($\ep_k=-1$ for fermionic field) probability.
Following the definition of $N^0$ in Eq.~(\ref{N_0}),
Eqs.~(\ref{N_qvkv_BB}) and (\ref{N_qvkv_FF}) can be simplified as
\be
N^0=-I^e_\pi (4\vk^2)~~~ {\rm for~}\pi\pi~{\rm loop}~,
\label{N0_pi}
\ee
and
\be
N^0=-I^e_N (8\vk^2)~~~ {\rm for~}NN~{\rm loop}~.
\label{N0_N}
\ee
Using the above Eqs.~(\ref{N0_pi}) and (\ref{N0_N}) in Eq.~(\ref{el_last})
as well as their relevant parameters from Eq.~(\ref{pi_case}) and (\ref{N_case}),
we get the electrical conductivity of the pionic
and nucleonic medium:
\be
\sigma_\pi=\frac{\beta e^2}{3}\int^{\infty}_{0} 
\frac{d^3\vk}{(2\pi)^3}\frac{\vk^2}{{\om^\pi_k}^2\Gamma_\pi}n_k(\om^\pi_k)
\{1+n_k(\om_k^\pi)\} 
\label{el_pi}
\ee
and
\bea
\sigma_N&=&\frac{2\beta e^2}{3}\int^{\infty}_{0} 
\frac{d^3\vk}{(2\pi)^3}\frac{\vk^2}{{\om^N_k}^2\Gamma_N}[n^+_k(\om_k^N)\{1-n^+_k(\om^N_k)\}
\nn\\
&&+n^-_k(\om^N_k)\{1-n^-_k(\om^N_k)\}]~.
\label{el_N}
\eea
Hence, adding the pionic and nucleonic components, we get the total electrical conductivity
\be
\sigma_{\rm T}=\sigma_{\pi}+\sigma_N~.
\ee

\section{Thermal width}
\label{sec:width}
Let us come to the thermal widths of pion ($\Gamma_\pi$) and nucleon 
($\Gamma_N$). Pion thermal width can be obtained from the imaginary part
of pion self-energy for different mesonic and baryonic fluctuations.
Fig.~\ref{El_pi_N}(b) represents pion self-energy diagram for
$\pi M$ (mesonic) loops - ${\Pi}^R_{\pi(\pi M)}$, where $M=\sigma,~ \rho$.
Here subscript in ${\Pi}^R_{\pi(\pi M)}$ stands for the 
external (outside the bracket) and internal (inside the bracket) 
particles for the diagram~\ref{El_pi_N}(b). This notation will be followed
by latter diagrams also. Now, pion self-energy for different 
baryonic loops (${\Pi}^R_{\pi(NB)}$) can have two possible diagrams 
as shown in Fig.~\ref{El_pi_N}(c) and (d). Here internal lines $NB$
stand for nucleon ($N$) and baryon ($B$) respectively, where different
4-star spin $1/2$ and $3/2$ baryons are taken in our calculations. Adding 
all those mesonic and baryonic loops, we get total thermal width of pion
$\Gamma_\pi$, which can be expressed as
\bea
\Gamma_\pi&=& \sum_M\Gamma_{\pi(\pi M)} + \sum_B\Gamma_{\pi(NB)}
\nn\\
&=&-\sum_M{\rm Im}{\Pi}^R_{\pi(\pi M)}(k_0=\om^\pi_k,\vk)/m_\pi 
\nn\\
&&~~~-\sum_B{\rm Im}{\Pi}^R_{\pi(NB)}(k_0=\om^\pi_k,\vk)/m_\pi~.
\label{Gam_pi}
\eea
Similarly, nucleon self-energy is shown in Fig.~\ref{El_N_piB}(b)
and it has been denoted as $\Sigma^R_{N(\pi B)}$, where in internal
lines, we have taken all those spin $1/2$ and $3/2$ baryons ($B$) as
taken in pion self-energy for baryonic loops. Hence, summing these all
$\pi B$ loops, we can express our nucleon thermal width as 
\be
\Gamma_N=\sum_B\Gamma_{N(\pi B)}=-\sum_B{\rm Im}\Sigma^R_{N(\pi B)}(k_0=\om^N_k,\vk)~.
\label{Gam_N}
\ee

Next we discuss briefly the calculations of thermal widths from different one-loop self-energy
graphs as shown in Fig.~(\ref{El_pi_N}) and (\ref{El_N_piB}).

\subsection{Pion thermal width for different mesonic loops}
\label{subsec:pi_mes}
To calculate the mesonic loop contribution of pionic thermal width
$\Gamma_{\pi(\pi M)}$, the pion self-energy for $\pi M$ loops, where $M$ stands for $\sigma$
and $\rho$ mesons, have been evaluated and it is expressed as~\cite{GKS}
\bea
\Gamma_{\pi(\pi M)}&=&{\rm Im}{\Pi}^R_{\pi(\pi M)}(k_0=\om^\pi_k,\vk)/m_\pi
\nn\\
&=&\frac{1}{m_\pi} \int \frac{d^3 \vl}{32\pi^2 \om_l^\pi\om_u^M}
\nn\\
&& L(l_0=-\om^\pi_l, \vl, k_0=\om^\pi_k, \vk) \{n(\om^\pi_l) 
\nn\\
&& -n(\om^M_u)\}\delta(\om^\pi_k +\om^\pi_l - \om^M_u)~,
\label{G_pi_piM}
\eea
where $n(\om^\pi_l)$, $n(\om^M_u)$ are BE distribution functions of $\pi$, $M$
mesons with energies $\om^\pi_l=(\vl^2+m_\pi^2)^{1/2}$ and 
$\om^M_u=(|\vk-\vl|^2+m_M^2)^{1/2}$ respectively.
The vertex factors $L(k,l)$~\cite{GKS}
have been obtained from the effective Lagrangian density,
\be
{\cal L} = g_\rho \, {\vec \rho}_\mu \cdot {\vec \pi} \times \del^\mu {\vec \pi} 
+ \frac{g_\sigma}{2} m_\sigma {\vec \pi}\cdot {\vec\pi}\,\sigma~.
\label{Lag_pipiM}
\ee

\subsection{Pion thermal width for different baryonic loops}
Along with the mesonic fluctuations, different baryon fluctuations
may provide some contributions in pion thermal width. This component
can be derived from pion self-energy for different $NB$ loops,
where $B =N(940)$, $\Delta(1232)$, $N^*(1440)$, $N^*(1520)$,
$N^*(1535)$, $\Delta^*(1600)$, $\Delta^*(1620)$, $N^*(1650)$, 
$\Delta^*(1700)$, $N^*(1700)$, $N^*(1710)$, $N^*(1720)$ are
taken~\cite{G_pi_JPG,G_eta_BJP}. The masses of all the 4-star 
baryon resonances (in MeV) are presented inside the brackets.
The direct and cross diagrams of pion self-energy 
for $NB$ loops are shown in Fig.~\ref{El_pi_N}(c) and (d).
Adding the relevant Landau cut contributions of both diagrams (c) 
and (d), the total thermal width of pion for any $NB$ loop is 
given by~\cite{G_pi_JPG,G_eta_BJP}
\bea
\Gamma_{\pi(NB)}&=& {\rm Im}{\Pi}^R_{\pi(NB)}(k_0=\om^\pi_k,\vk)/m_\pi
\nn\\
&=&\frac{1}{m_\pi} \int \frac{d^3 \vl}{32\pi^2 \om_l^N\om_u^B}
\nn\\
&& [L(l_0=\om^N_l,\vl, k_0=\om_k^\pi,\vk)\{n^+_l(\om^N_l)
\nn\\
&& - n^+_u(\om^B_u)\}\delta(\om^\pi_k -\om^N_l + \om^B_u)
\nn\\
&&+L(l_0=-\om^N_l,\vl, k_0=\om_k^\pi,\vk)\{-n^-_l(\om^N_l) 
\nn\\
&&+ n^-_u(\om^B_u)\}\delta(\om^\pi_k +\om^N_l - \om^B_u)]~,
\label{G_pi_NB}
\eea
where $n^{\pm}(\om^N_l)$, $n^{\pm}(\om^B_u)$ are FD distribution functions of 
$N$, $B$ ($\pm$ for particle and anti-particle)
with energies $\om^N_l=(\vl^2+m_N^2)^{1/2}$ and 
$\om^B_u=(|\pm\vk +\vl|^2+m_B^2)^{1/2}$ ($\pm$ for two different diagrams) 
respectively.

With the help of the effective Lagrangian densities~\cite{Leopold},
\bea
{\cal L}&=&\frac{f}{m_\pi}{\ov \psi}_B\gamma^\mu
\left\{
\begin{array}{c}
i\gamma^5 \\
\unit
\end{array}
\right\}
\psi_N\del_\mu\pi + {\rm h.c.}~{\rm for}~J_B^P=\frac{1}{2}^{\pm},
\nn\\
{\cal L}&=&\frac{f}{m_\pi}{\ov \psi}^\mu_B
\left\{
\begin{array}{c}
\unit \\
i\gamma^5
\end{array}
\right\}
\psi_N\del_\mu\pi + {\rm h.c.}~{\rm for}~J_B^P=\frac{3}{2}^{\pm},
\label{Lag_BNpi}
\eea
the vertex factors $L(k,l)$~\cite{G_pi_JPG,G_eta_BJP} 
can be found.

\subsection{Nucleon thermal width}
\label{subsec:N}
The nucleonic thermal width has been calculated from
nucleon self-energy for different possible $\pi B$ loops, 
where $B$ stands for all the baryons as taken in pion self-energy
for baryonic loops. Evaluating the loop diagram, shown in 
Fig.~\ref{El_N_piB}(b), we get~\cite{G_NNst_BJP,G_N}
\bea
\Gamma_{N(\pi B)} &=&-\sum_B{\rm Im}\Sigma^R_{N(\pi B)}(k_0=\om^N_k,\vk)
\nn\\
&=&\int \frac{d^3 \vl}{32\pi^2 \om_l^\pi\om_u^B}
\nn\\
&& L(l_0=-\om^\pi_l,\vl, k_0=\om_k^N,\vk)\{n(\om^\pi_l) 
\nn\\
&&+ n^+(\om^B_u)\} \delta(\om^N_k +\om^\pi_l - \om^B_u)
\label{G_N_piB}
\eea
where $n(\om^\pi_l)$ and $n^+(\om^B_u)$ are BE and 
FD distribution functions for $\pi$ and $B$ with energies
$\om_l^\pi=(\vl^2+m_\pi^2)^{1/2}$ and $\om^B_u=(|\vk - \vl|^2+m_B^2)^{1/2}$
respectively.

The vertex factors $L(k,l)$~\cite{G_NNst_BJP,G_N}
can be deduced by using the $\pi NB$ interaction Lagrangian 
densities from Eq.~(\ref{Lag_BNpi}).
%
%
%
\section{Results and Discussion}
\label{sec:num}
%
%
%
%
%
%
%
\begin{figure}
\begin{center}
\includegraphics[scale=0.35]{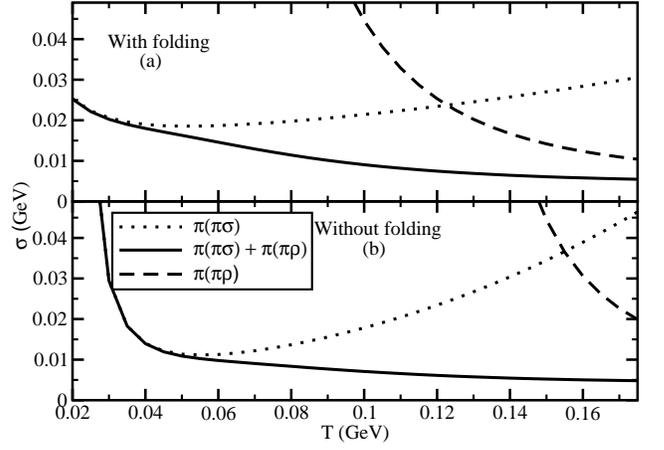}
\caption{Temperature dependence of electrical conductivity
pionic medium due to its different mesonic fluctuations - $\pi\sigma$
(dotted line), $\pi\rho$ (dashed line) loops and their total (solid line).
With and without folding effect of resonances $M=\sigma, \rho$ are taken
in upper and lower panels respectively.} 
\label{el_ppM_T}
\end{center}
\end{figure}
\begin{figure}
\begin{center}
\includegraphics[scale=0.35]{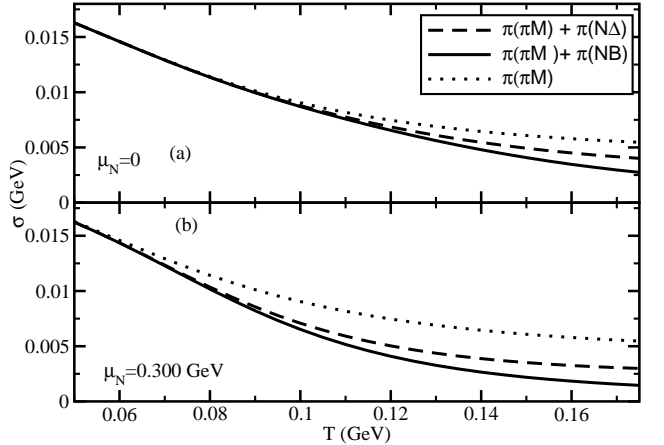}
\caption{Effect of baryonic fluctuations ($N\Delta$ loop : dashed line,
$NB$ loops : solid line) after adding with mesonic fluctuations 
($\pi M$ loops : dotted line) of pion on $\sigma_\pi(T)$ at
$\mu_N=0$ (a) and $\mu_N=0.300$ GeV (b).} 
\label{el_pNB_T}
\end{center}
\end{figure}
\begin{figure}
\begin{center}
\includegraphics[scale=0.35]{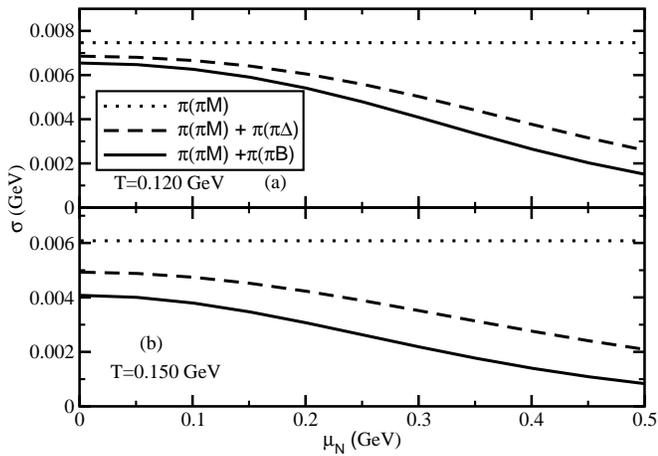}
\caption{Same as Fig.~(\ref{el_pNB_T}) for $\sigma_\pi(\mu_N)$ at
$T=0.120$ GeV (a) and $T=0.150$ GeV (b).} 
\label{el_pNB_mu}
\end{center}
\end{figure}
\begin{figure}
\begin{center}
\includegraphics[scale=0.35]{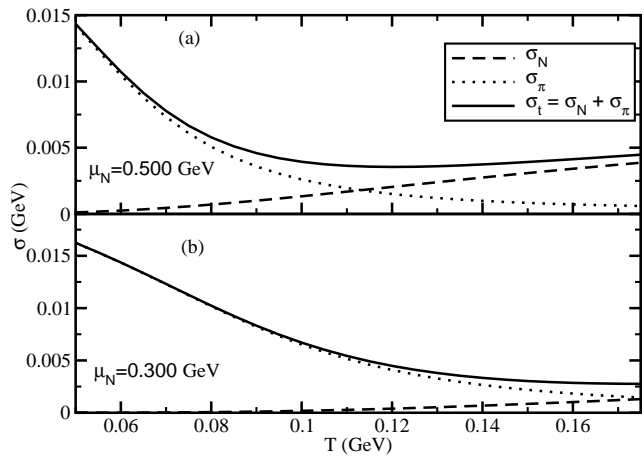}
\caption{Temperature dependence of electrical conductivity for 
pion (dotted line), nucleon (dashed line) components and their
total (solid line) at $\mu_N=0.500$ GeV (a) and $\mu_N=0.300$ GeV.} 
\label{el_pNt_T}
\end{center}
\end{figure}
\begin{figure}
\begin{center}
\includegraphics[scale=0.35]{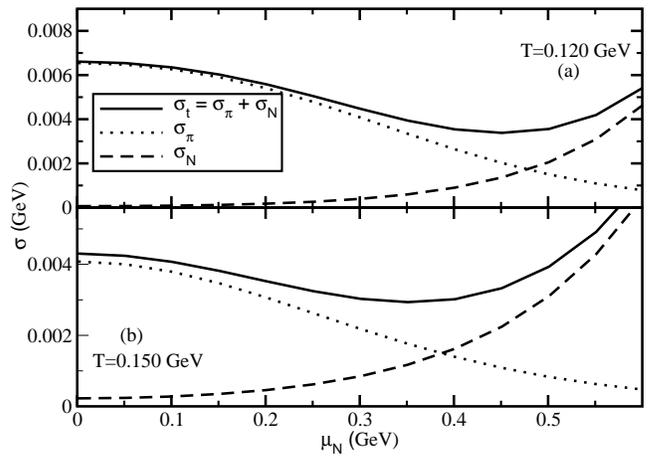}
\caption{$\mu_N$ dependence of electrical conductivity for 
pion (dotted line), nucleon (dashed line) components and their
total (solid line) at $T=0.120$ GeV (a) and $T=0.150$ GeV.} 
\label{el_pNt_mu}
\end{center}
\end{figure}
Using the $\Gamma_{\pi(\pi\sigma)}(\vk,T)$, $\Gamma_{\pi(\pi\rho)}(\vk,T)$ and
their total in the integrand of Eq.~(\ref{el_pi}), the dotted, dashed
and solid lines of Fig.~(\ref{el_ppM_T}) are generated, where folding~\cite{GKS} 
by vacuum spectral functions of resonances $\sigma$ and $\rho$ are considered 
in panel (a) but not in panel (b). Like the results of shear viscosity in
the earlier work~\cite{GKS}, $\sigma$ and $\rho$ resonances play dominant role in
the electrical conductivity at low ($T<0.100$ GeV) and high ($T>0.100$ GeV) 
temperature domain respectively. We get $\sigma_\pi(T)$ as a decreasing function
in low and high temperature both, although a mild increasing function of shear
viscosity $\eta_\pi(T)$ has been observed in Ref.~\cite{GKS} at high temperature 
domain of hadronic matter ($0.100$ GeV $<T<0.175$ GeV). The mathematical origin
for this differences in the nature of $\sigma_\pi(T)$ and $\eta_\pi(T)$ is
because of different power of momentum ($\vk^4$ for $\sigma_\pi$ but $\vk^6$ 
for $\eta_\pi$) in the numerator of their respective integrand.

Adding baryonic loop contributions with the mesonic loops of pion self-energy,
we get total thermal width of pion as described explicitly in Eq.~(\ref{Gam_pi}).
Fig.~\ref{el_pNB_T}(a) and (b) for $\mu_N=0$ and $0.300$ GeV reveal that $\sigma_\pi(T)$
reduces after adding baryonic loop contribution in pion self-energy and its reduction
strength becomes larger for larger values of $\mu_N$ as baryonic loop contribution, 
$\Gamma_{\pi(NB)}(\vk,T,\mu_N)$ depends sensitively on $\mu_N$. To display the 
dominant contribution of $N\Delta$ loop ($\Gamma_{\pi(N\Delta)}$), 
Fig.~(\ref{el_pNB_T}) shows individual contributions of meson loops,
meson loops + $N\Delta$ loop and meson + baryon loops by dotted,
dashed and solid lines respectively.

Next, Fig.~\ref{el_pNB_T}(a) and (b) for $T=0.120$ GeV and $0.150$ GeV
show $\mu_N$ dependence of electrical conductivity of pionic component
for meson loops (dotted line), meson loops + $N\Delta$ loop (dashed line) 
and meson + baryon loops (solid line). As $\Gamma_{\pi(\pi M)}(\vk,T)$ 
is independent of $\mu_N$, therefore corresponding $\sigma_\pi$ 
(dotted line) remain constant with the variation of $\mu_N$. After
adding $N\Delta$ loop (dashed line) and other baryon loops (solid line),
a decreasing nature of $\sigma_\pi(\mu_N)$ are clearly noticed. A sensitive
dependence of $\mu_N$ in $\Gamma_{\pi(NB)}$ for $N\Delta$ loop (dominant)
and other baryon loops are the main reason behind the decreasing nature
of $\sigma_\pi(\mu_N)$.
\begin{figure}
\begin{center}
\includegraphics[scale=0.8]{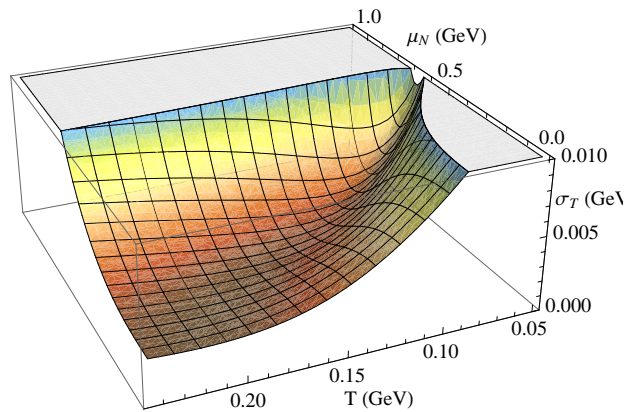}
\caption{(Color online) Total electrical conductivity $\sigma_T$ in $T$-$\mu_N$ plane.} 
\label{El_muT}
\end{center}
\end{figure}
\begin{figure}
\begin{center}
\includegraphics[scale=0.35]{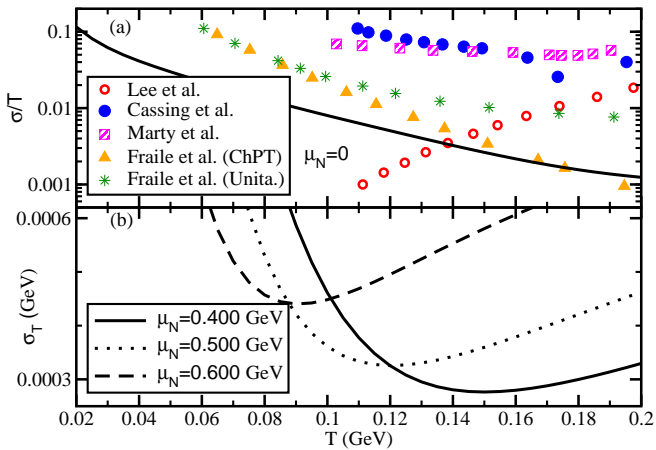}
\caption{(Color online) Our results of $\sigma(T,\mu_N=0)/T$
are compared with the results of Refs.~\cite{Cassing,Marty,Lee,Nicola_PRD} (a).
Valley structure of $\sigma(T)$ at $\mu_N=0.400$ GeV, $0.500$ GeV, $0.600$ GeV
are shown by solid, dotted and dashed lines respectively in panel (b).} 
\label{comp}
\end{center}
\end{figure}
\begin{figure}
\begin{center}
\includegraphics[scale=0.35]{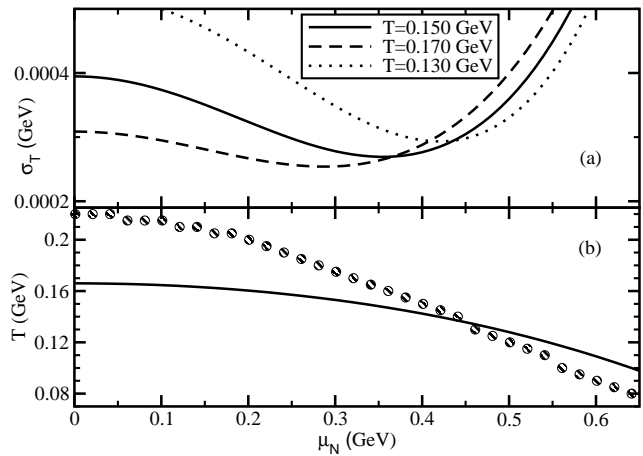}
\caption{(a): Valley structure of $\sigma(\mu_N)$ at three different 
$T$. (b): The points of minima (solid circles) and freeze 
out line~\cite{freezeout} (solid line) are shown in $T$-$\mu_N$ plane.} 
\label{el_t_mu2}
\end{center}
\end{figure}

In Fig.~\ref{el_pNt_T}(a) and (b) for $\mu_N=0.500$ GeV and $0.300$ GeV,
the $T$ dependence of pionic ($\sigma_\pi$), nucleonic ($\sigma_N$) components of 
electrical conductivities and their total ($\sigma_T$) are shown by
dotted, dashed and solid lines respectively.
Corresponding results in $\mu_N$ axis are shown in 
Fig.~\ref{el_pNt_mu}(a) and (b) for $T=0.120$ GeV and $0.150$ GeV.
Unlike to $\sigma_\pi$, the $\sigma_N$ increases with both $T$ and $\mu_N$.
The nucleon phase space factors or statistical weight
factors of FD distributions in $\sigma_N$ are playing a dominant over 
the $\Gamma_N(T, \mu_N)$, whereas for pionic case, $\Gamma_\pi(T,\mu_N)$ becomes more
influential than pionic phase factors or statistical Bose enhanced 
weight factors in $\sigma_\pi$. This is the mathematical reason for opposite
nature of $\sigma_\pi(T,\mu_N)$ and $\sigma_N(T,\mu_N)$. From a simultaneous
observation of Fig.~(\ref{el_pNt_T}) and (\ref{el_pNt_mu}), we can conclude
that the decreasing nature of $\sigma_T(T,\mu_N)$ becomes inverse beyond
a certain points of $T$ and $\mu_N$, where $\sigma_T$ exposes the points of minima.
This behavior can be visualized well from Fig.~(\ref{El_muT}), which
exhibits 3-dimensional plot of $\sigma_T(T,\mu_N)$.

Up to now, our results are presented as normalized values of $e^2$
(in other word we have taken $e^2=1$) but exact values of $\sigma_T$
(after multiplying by $e^2=4\pi/137$) have been shown in the last 
two figures (\ref{comp}) and (\ref{el_t_mu2}).
Fig.~\ref{comp}(a) displays a comparison of present results
with the earlier results, obtained by Fraile et al.~\cite{Nicola}
(stars and triangles), Lee et al.~\cite{Lee} (open circles), Marty
et al.~\cite{Marty} (squares), Cassing et al.~\cite{Cassing} 
(solid circles) at hadronic temperature domain for $\mu_N=0$.
Within $0.110$ GeV $<T<0.175$ GeV, present results more or less
agrees with the results of Ref.~\cite{Nicola,Lee} but quite smaller
than the results of Ref.~\cite{Marty,Cassing}. Fig.~\ref{comp}(b)
shows $\sigma_T$ vs $T$ at three different values of $\mu_N$, where
we notice the shifting of minimum values of $\sigma_T$ towards 
lower $T$ as one increases $\mu_N$. 
Alternatively, these minimum
values of $\sigma_T$ will also be shifted towards lower $\mu_N$ as $T$
will increase, which is explicitly shown in Fig.~\ref{el_t_mu2}(a).
Next, Fig.~\ref{el_t_mu2}(b) represents the points of minima for $\sigma_T$
in $T$-$\mu_N$ plane. An approximated freeze out line (solid line),
taken from Ref.~\cite{freezeout}, is also pasted in Fig.~\ref{el_t_mu2}(b).
The points of minima, which are located outside the freeze out line, can only
be covered by the expanding fireball, produced in different beam energies
of heavy ion collisions. Therefore, the minima or valley structure can be observed
from ($T_f\approx 0.166$ GeV, $\mu_f\approx 0$) to ($T_f\approx 0.140$ GeV,
$\mu_f\approx 0.420$ GeV), where subscript $f$ stands for freeze out. 
In other word, from high beam energy $\sqrt{s}$
like RHIC experiment ($\sqrt{s}=200$ GeV), this valley structure can be
observed up to $\sqrt{s}\approx 8$ GeV. However, within this long range of beam 
energy or freeze out line, some points of minima may cross the quark-hadron
transition line and therefore, they may not be observed in the experiment.
One should keep in mind that these minima or valley structure is completely
appeared due to phase space effect of hadronic medium and has nothing relation
with the quark-hadron transition. Therefore, one can observe only those points
of minima of hadronic medium, which will be in between freeze out and quark-hadron
transition lines. Although, there is some possibility for not observing any
points of minima, if they all are located in quark phase domain of $T$-$\mu_N$ plane.
In this regards, we can say at least that $\sigma(T,\mu_N)$ of hadronic medium
decreases as one goes towards quark-hadron transition line.

We have presented the numerical values of $\sigma(T,\mu_N=0)/T$,
estimated by earlier works in Table~\ref{tab}, where most of the works
are displaying the decreasing $\sigma(T)/T$ in hadronic 
temperature~~\cite{Cassing,Marty,Nicola_PRD,Nicola} 
and increasing $\sigma(T)/T$ in temperature domain of
quark phase~\cite{Cassing,Marty,Puglisi,Greif,PKS,Finazzo,LQCD_Amato}. 
Among them, Refs.~\cite{Cassing,Marty}, covering the both temperature domain, 
have exhibited the minimum value of $\sigma/T$ near transition temperature.
On the basis of these earlier results at $\mu_N=0$ and our estimations
at finite $\mu_N$ within the $T$-$\mu_N$ domain of hadronic matter, a valley
structure along quark-hadron transition line in $T$-$\mu_N$ plane may be expected
and this issue may be confirmed after further research on $\sigma$-calculations at finite
baryon density, based on different effective QCD model.
\begin{table}  
\begin{center}
\label{tab}
\begin{tabular}{|c|c|c|}
\hline
 & $\sigma/T$ at  & $\sigma/T$ at  \\
 &  $T=(0.120$  &  $T=(0.175$ \\
& -$~0.175)$ GeV & -$~350)$ GeV  \\
\hline
\hline
\underline{LQCD Results:} & & \\
Gupta~\cite{LQCD_Gupta} & - & $\approx0.375$ \\
Ding et al.~\cite{LQCD_Ding} & - & $\approx0.033(+0.018,$ \\
& & $~~~~~~-0.016)$ \\
Arts et al.~\cite{LQCD_Arts_2007} & - & $\approx0.020(\pm0.005)$ \\
Brandt et al.~\cite{LQCD_Barndt} & - & $\approx0.020(\pm 0.006)$ \\
Burnier et al.~\cite{LQCD_Burnier} & - & $\approx0.0064$ \\
Amato et al.~\cite{LQCD_Amato} & - & $\approx0.003(\pm 0.001)$ \\
& & -$~0.015(\pm 0.003)$ \\
Buividovich et al.~\cite{LQCD_Buividovich} & - & $\approx0.0021(\pm 0.0003)$ \\
\hline
Yin~\cite{Yin} & - & $\approx0.06(+0.04,$\\
 & & $-0.02)$  \\
\hline
Puglisi et al.~\cite{Puglisi} & - & $\approx0.09~$-$~0.13$ \\
(PQCD in RTA) & & \\
Puglisi et al.~\cite{Puglisi} & - & $\approx0.01~$-$~0.07$ \\
(QP in RTA) & & \\
\hline
Greif et al.~\cite{Greif} & - & $\approx0.04~$-$~0.06$ \\
(BAMPS) & & \\
\hline
Marty et al.~\cite{Marty} & - & $\approx0.06~$-$~0.16$ \\
(DQPM) & & \\
Marty et al.~\cite{Marty} & $\approx0.06~$-$~0.05$ & $\approx0.05~$-$~0.5$ \\
(NJL) & & \\
\hline
Cassing et al.~\cite{Cassing} & $\approx0.088~$-$~0.025$ & $\approx0.025~$-$~0.2$ \\
(PHSD) & & \\
\hline
Finazzo et al.~\cite{Finazzo} & $\approx0.004~$-$~0.010$ & $\approx0.010~$-$~0.015$ \\
\hline
Lee et al.~\cite{Lee} & $\approx0.001~$-$~0.011$ & $\approx0.36~$-$~0.015$ \\
\hline
Fraile et al.~\cite{Nicola} & $\approx0.013~$-$~0.010$ & - \\
(Unitarization) & & \\
Fraile et al.~\cite{Nicola} & $\approx0.008~$-$~0.002$ & - \\
(ChPT) & & \\
\hline
Present Results & $\approx0.004~$-$~0.001$ & - \\
\hline
\end{tabular}
\caption{At $\mu_N=0$, the $\sigma(T)/T$ in approximated temperature
domain of hadronic ($T\approx 0.120$ GeV to $0.175$ GeV) and quark 
($T\approx 0.175$ GeV to $0.350$ GeV) phases are presented in 
2nd and 3rd columns, whereas in 1st column, the references 
(with their methodologies) are addressed.}
\end{center}
\end{table}
%


\section{Summary and Conclusion}
\label{sec:concl}
The present work provide an estimation of electrical conductivity
of hadronic medium at finite temperature and baryon density. Assuming
pion and nucleon as most abundant medium constituents, we have first
deduced thermal correlators of their electromagnetic currents and then,
taking the static limit of these correlators, the expressions
of electrical conductivities for pionic and nucleonic components are derived.
For getting the non divergent values of these correlators in the static limit,
one has to include the finite thermal widths of the medium constituents - pion
and nucleon. This is a traditional quasi-particle technique of Kubo frame work,
used during the calculations of transport coefficients from the relevant correlators 
in their static limits. Following the field theoretical version of optical theorem,
the thermal widths of pion and nucleon are obtained from the imaginary part
of their one-loop self-energy diagrams, which accommodate different mesonic
and baryonic resonances in the intermediate states. As a dynamical part, 
the interaction of pion and nucleon with other mesonic and baryonic resonances 
are guided by the effective hadronic Lagrangian densities, where their couplings
are tuned by the decay width of resonances, based on the experimental data from PDG. 
The momentum distribution of
these thermal widths are integrated out during evaluation of electrical conductivities
of respective components. 

The electrical conductivity for pionic component is obtained as a decreasing function
$T$ and $\mu_N$, where mesonic loops are dominant to fix its numerical strength.
The $\pi\sigma$ and $\pi\rho$ loops of pion self-energy control the strength
of electrical conductivity at low and high $T$ regions respectively. While a further
reduction of numerical values in conductivity at high $T$ domain is noticed after
addition of different baryonic loops in pion self-energy. Electrical conductivity
of pionic component due to mesonic loops remain constant with $\mu_N$ but it is 
transformed to a decreasing function when the baryonic loops are added in the pion
self-energy. The nucleonic component give the increasing values of electrical
conductivity with the variation of $T$ and $\mu_N$. After adding these pionic
and nucleonic components, the total electrical conductivity first decreases
at pion dominating $T$-$\mu_N$ domain and then increases at nucleonic dominating
domain. Therefore, the numerical results show a set of $T$-$\mu_N$ points, 
where total electrical conductivity becomes minimum and this valley structure
in $T$-$\mu_N$ plane can only be observed if the points of minima are located
between freeze out line and quark-hadronic transition line. 

Comparing with earlier
estimations of electrical conductivity at $\mu_N=0$,
present work more or less agrees with Refs.~\cite{Nicola,Lee} quantitatively 
and quantitatively it is similar with most of the earlier 
works~\cite{Cassing,Marty,Nicola_PRD,Nicola,Greif2}, which show that electrical
conductivity at $\mu_N=0$ decreases with $T$. On the basis of these
earlier studies at $\mu_N=0$ and present investigation at finite $\mu_N$,
a general decreasing nature in the numerical values of electrical conductivity
for hadronic matter is observed when one goes from freeze out to quark-hadron
transition line in $T$-$\mu_N$ plane. Further research in different model
calculations at finite $\mu_N$ may confirm this conclusion.

{\bf Acknowledgment :}
The work is financially supported from UGC Dr. D. S. Kothari Post Doctoral Fellowship under
grant No. F.4-2/2006 (BSR)/PH/15-16/0060.
\section{Appendices}
\subsection{Calculation $N(\vq,\vk)$}
\label{cal_Nqk}
Let us write the 11-component of two point function of current-current 
correlator in terms of field operators. For $\phi_\pi$ field
it is given by
\bea
\Pi_{11}(q)&=&i\int d^4x e^{iqx}\langle TJ^{\rm EM}_{\mu}(x) J_{\rm EM}^{\mu}(0) \rangle_\beta
\nn\\
&=&ie^2\int d^4x e^{iqx}\langle T \phi_\pi(x)\del_\mu\phi_\pi(x)
\phi_\pi(0)\del^\mu\phi_\pi(0)\rangle_\beta~.
\nn\\
\label{pi_ab_Ap}
\eea
With the help of the Wick's contraction technique, we have
\bea
\Pi_{11}(q)&=&ie^2\int d^4x e^{iqx}
[\langle T \phi_\pi\underbrace{(x)\del_\mu\phi_\pi\overbrace{(x)
\phi_\pi}(0)\del^\mu\phi_\pi}(0)\rangle_\beta
\nn\\
&=&ie^2\int \frac{d^4k}{(2\pi)^4}N(q,k) 
D_{11}(k)D_{11}(p=q-k)~,
\nn\\
\eea
where 
\be
N(q,k)= (-4)k^\mu(q-k)_\mu
\ee
and its space component part is
\be
N(\vq,\vk)= (-4)\{-\vk\cdot\vq + \vk^2\}~.
\ee

Similarly for $\psi_N$ field,
\bea
\Pi_{11}(q)&=&ie^2\int d^4x e^{iqx}
\langle T{\ov \psi}_N\underbrace{(x)\gamma_\mu\psi_N\overbrace{(x)
{\ov \psi}_N}(0)\gamma^\mu\psi_N}(0)\rangle_\beta
\nn\\
&=&ie^2\int \frac{d^4k}{(2\pi)^4}N(q,k) 
D_{11}(k)D_{11}(p=q+k)~,
\nn\\
\eea
where 
\bea
N(q,k)&=&{\rm Tr}[\gamma^\mu(\qs+\ks+m_\psi)
\gamma_\mu(\ks+m_\psi)]
\nn\\
&=&8k^\mu(q+k)_\mu -4[k\cdot(q+k)-m_\psi^2]g^\mu_\mu
\eea
and the space component part of 
\be
N(q,k_0=\pm \om_k,\vk)=8k^\mu(q+k)_\mu -4[k\cdot q]g^\mu_\mu
\ee
is
\be
N(\vq,\vk)=-8\vk\cdot(\vq +\vk) +4[\vk\cdot \vq]g^i_i~.
\ee
\subsection{Application of L'Hospital rule}
\label{LHos}
For finite value of $\Gamma$, the Eq.~(\ref{el_G}) becomes
\be
\sigma=\frac{e^2}{3}\int\frac{d^3k}{(2\pi)^3}\frac{N^0}{4\om_k^2\Gamma}
\lim_{q_0,\vq \rightarrow 0}\left[\frac{C_2}{q_0}
+\frac{C_3}{q_0}\right]~,
\label{el_G_finite}
\ee
as
\be
\lim_{\vq \rightarrow 0}\om_p=\om_k~.
\ee
Applying L'Hospital's rule, we can write
\bea
\lim_{q_0 \rightarrow 0}\frac{C_{2,3}(q_0)}{q_0}&=&\lim_{q_0\rightarrow 0}
\frac{\frac{d}{dq_0}\{C_{2,3}(q_0)\}}{\frac{d}{dq_0}\{q_0\}}
\nn\\
&=&\frac{d}{dq_0}\{ \pm n^{\mp}_p(\om_p=\mp q_0+\om_k) \}
\nn\\
&=&\beta [n^{\mp}_k(1 +\ep_k n^{\mp}_k)]~,
\eea
since
\bea
\left(\pm \right)\frac{d}{dq_0}n^{\mp}_p(\om_p=\mp q_0+\om_k)&=&\left(\pm\right)
\frac{-\beta \frac{d\om_p}{dq_0} 
e^{\beta(\om_p \pm \mu)}}{\{e^{\beta(\om_p \pm \mu)}+\ep_k\}^2}
\nn\\
\lim_{q_0\rightarrow 0}\left(\pm \right)\frac{d}{dq_0}n^{\mp}_p(\om_p=\mp q_0+\om_k)
&=&\left(\pm\right)\frac{-\beta \left(\mp \right)
 e^{\beta(\om_k \pm \mu)}}{\{e^{\beta(\om_k \pm \mu)}+\ep_k\}^2}
\nn\\
&=&\beta [n^{\mp}_k(1 +\ep_k n^{\mp}_k)]~.
\nn\\
\eea

%
%
%

\end{document}